\documentclass[12pt,aps,amsmath,latexsym,amsfonts,letterpaper]{JHEP3}
\usepackage[all]{xy}
\usepackage{epsfig}
\usepackage{subfigure}

\def\half{\textstyle{1\over2}}

\def\ie{{\it i.e.,}\ }

\font\mybb=msbm10 at 12pt 
\def\bb#1{\hbox{\mybb#1}}
\def\bZ {\bb{Z}}

\newcommand{\be}{\begin{equation}}
\newcommand{\ee}{\end{equation}}
\newcommand{\bea}{\begin{eqnarray}}
\newcommand{\eea}{\end{eqnarray}}
\newcommand{\bml}{\begin{mathletters}}
\newcommand{\eml}{\end{mathletters}}

\newcommand{\la}{\ensuremath{\lambda}}

\newcommand{\del}{\ensuremath{\partial}}

\newcommand{\ba}{\begin{eqnarray}}
\newcommand{\ea}{\end{eqnarray}}

\newcommand{\lab}{\label}

\newcommand{\beq}{\begin{equation}}
\newcommand{\eeq}{\end{equation}}
\newcommand{\beqa}{\begin{eqnarray}}
\newcommand{\eeqa}{\end{eqnarray}}
\newcommand{\beqar}{\begin{eqnarray*}}
\newcommand{\eeqar}{\end{eqnarray*}}

\title{Decay of an inhomogeneous state via resonant tunnelling}
\author{Paul M. Saffin, Antonio Padilla and Edmund J. Copeland\\
School of Physics and Astronomy, University Park, University of
Nottingham, Nottingham NG7 2RD, UK}

\date{\today}

\maketitle

\abstract{We recently investigated the nature of resonant tunnelling in standard scalar
Quantum Field Theory, uncovering the conditions required for resonance. It was shown
that whereas the homogeneous false vacuum may decay via bubble nucleation, it may not
decay in a resonant fashion. The no-go theorem given there is circumvented in this
study by considering an initial state other than the homogeneous
false vacuum, and we confirm our mechanism by showing in an explicit model
how resonant tunnelling occurs. Using this model we demonstrate how the tunnelling rate
depends on the energy of specially constructed initial states, with these states corresponding to 
contracting spherical bubbles of some vacuum
that evolve to a minimum radius and then tunnel to another vacuum, instead of the classical
motion where the bubble would just start to expand.}


\begin{document}

\section{Introduction} 
\lab{sec:intro}
A now standard calculation in quantum mechanics is the tunnelling rate for a particle travelling towards
a potential barrier, resulting in an exponentially suppressed transmission rate if the energy of the
particle is lower than the height of the barrier~\cite{Landau:1977}.
One of the remarkable features of quantum mechanics
as a wave theory is that by adding another barrier one may actually {\it increase} the transmission rate
for certain values of the particle's energy~\cite{Merz}. This is in analogy with the Fabry-Perot interferometer, where 
two partially-silvered mirrors have a higher transmission rate, for light of specific wavelengths, than a single partially-silvered
mirror~\cite{Hernandez:1986}. 
The reason for this increased tunnelling rate, 
known as resonant tunnelling, can be traced to the existence of a bound state living between the two
barriers, and the tunnelling rate increases for those particles whose energies match the energy, or
energies, of these bound states. The possibility of increased tunnelling rates in field theory, and
indeed string theory, may be relevant in the string landscape, as noted by Tye~\cite{Tye:2006tg}, 
who suggested these may
lead to an efficient way of navigating the string landscape of vacua.

Resonant tunnelling in quantum mechanics also relies on the  existence of an intermediate bound state which is used as a springboard for the tunnelling amplitude. In~\cite{Copeland:2007qf}, we identified five properties that the analogous state in standard scalar quantum field theory must satisfy, and proved that no such state could exist. The implication is that the homogeneous false vacuum {\it cannot} decay via resonant tunnelling in such a theory. Whilst this no-go theorem cast some serious doubts on the relevance of resonant tunnelling to the string landscape, one could also use it as a guidebook, helping us to look in the right places for resonant tunnelling in quantum field theory.

To illustrate this point in a little more detail, recall the five well motivated conditions that the intermediate bound state was required to satisfy: (i) it should be a solution to the classical field equations, (ii) it should have zero energy relative to the homogeneous false vacuum, (iii) it should asymptote to the false vacuum, (iv) it should be stationary everywhere on two separate occasions and (v) it should satisfy a certain WKB quantization condition. In our discussion in~\cite{Copeland:2007qf}, we speculated that resonant tunnelling might be possible in other quantum field theories, or if one could justifiably relax one of these five conditions.

In this paper, we do the latter, noting that the five conditions are only relevant to the decay of a {\it homogeneous} region false vacuum. Of course, this is the standard approach to studying false vacuum decay~\cite{Coleman:1977py}, even though our Universe is never really in a homogeneous vacuum state. We therefore  consider the possibility that our initial state is {\it inhomogeneous}, but still asymptotically false vacuum. Now we can no longer justify imposing the ``zero energy'' condition, (ii), on the bound state in its current form. It should be replaced with the following statement of energy conservation: the intermediate bound state should have the same energy as the initial inhomogeneous state. We still define energy relative to the homogeneous false vacuum, but we now see that the bound state energy may be non-zero. This means our no-go theorem no longer applies and one might hope to find resonance. 

We will explicitly demonstrate resonant tunnelling from an inhomogeneous initial state. It turns out that the initial state must have certain properties in order to make it susceptible to resonant decay. In the example we will give, the initial state will correspond to a contracting bubble, with some minimum radius. At the stationary point, this state decays via resonant tunnelling to an expanding bubble of different vacuum. The mechanism for enhancing the tunnelling amplitude is precisely that outlined in~\cite{Copeland:2007qf}, with oscillons playing the role of the bound state. 

The rest of this paper is organised as follows: we begin by giving a brief introduction to resonant tunnelling in quantum mechanics, outlining the important concepts that will
be needed for the field theory generalisation. We shall then describe how the phenomenon is expected to appear
in field theory, before presenting a particular model which allows for a direct calculation. In section~\ref{sec:thinWall}, the results are extended to more general cases, at least  in the thin wall approximation, and we argue that resonant decay is by no means restricted to the specific example given in section~\ref{model}. Finally, we discuss our results in section~\ref{sec:conc}, and speculate as to how they may impact on the string landscape.

\section{Review of resonant tunnelling in quantum mechanics.} 
\lab{sec:review}

The simplest way to view resonant tunnelling is with the semi-classical approximation in the path
integral formalism\cite{Zhota:1990}. To set the scene, we consider the motion of a quantum particle
as it approaches a set of barriers given by Fig. \ref{fig:qmPot1}, such that the energy of the particle, $E$,
is lower than the height of either barrier. In the semi-classical approximation the paths which dominate
the path integral are those which obey the classical equation of motion in regions where $V<E$, and those
which obey a Wick-rotated version of the equation if $E<V$. In \cite{Banks:1973ps} the under-barrier regions ($E<V$)
were termed most probable escape paths (MPEPs), and for the full trajectory to make sense the classical paths
and MPEPs must be joined at turning points, i.e. the particle must be stationary there so that we avoid imaginary momentum
in the classical regions.

\FIGURE{\centerline{\includegraphics[width=8cm]{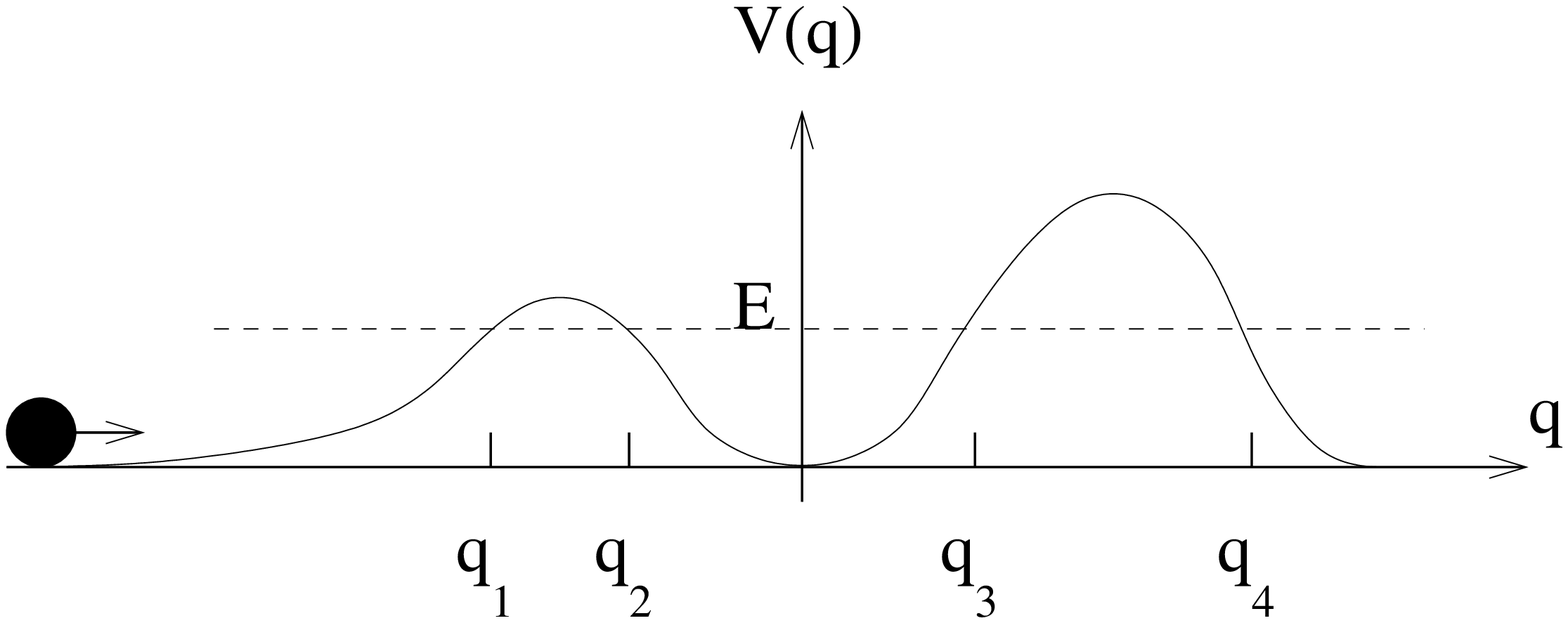}}
\caption{A quantum mechanics potential for resonant transmission.} \label{fig:qmPot1}}

Now that we know how to solve for the motion along a path we can construct a set of paths which
dominate the path integral, one such set is given in Fig. \ref{fig:resPhase} where we
show the five different regions that are relevant for the particle's motion: regions
I, III, V are the classically allowed regions of the potential in Fig. \ref{fig:qmPot1}, in those
the particle obeys the classical equations of motion; regions II and IV are classically forbidden, and in those places the
particle follows a MPEP.

\FIGURE{\centerline{\includegraphics[width=8cm]{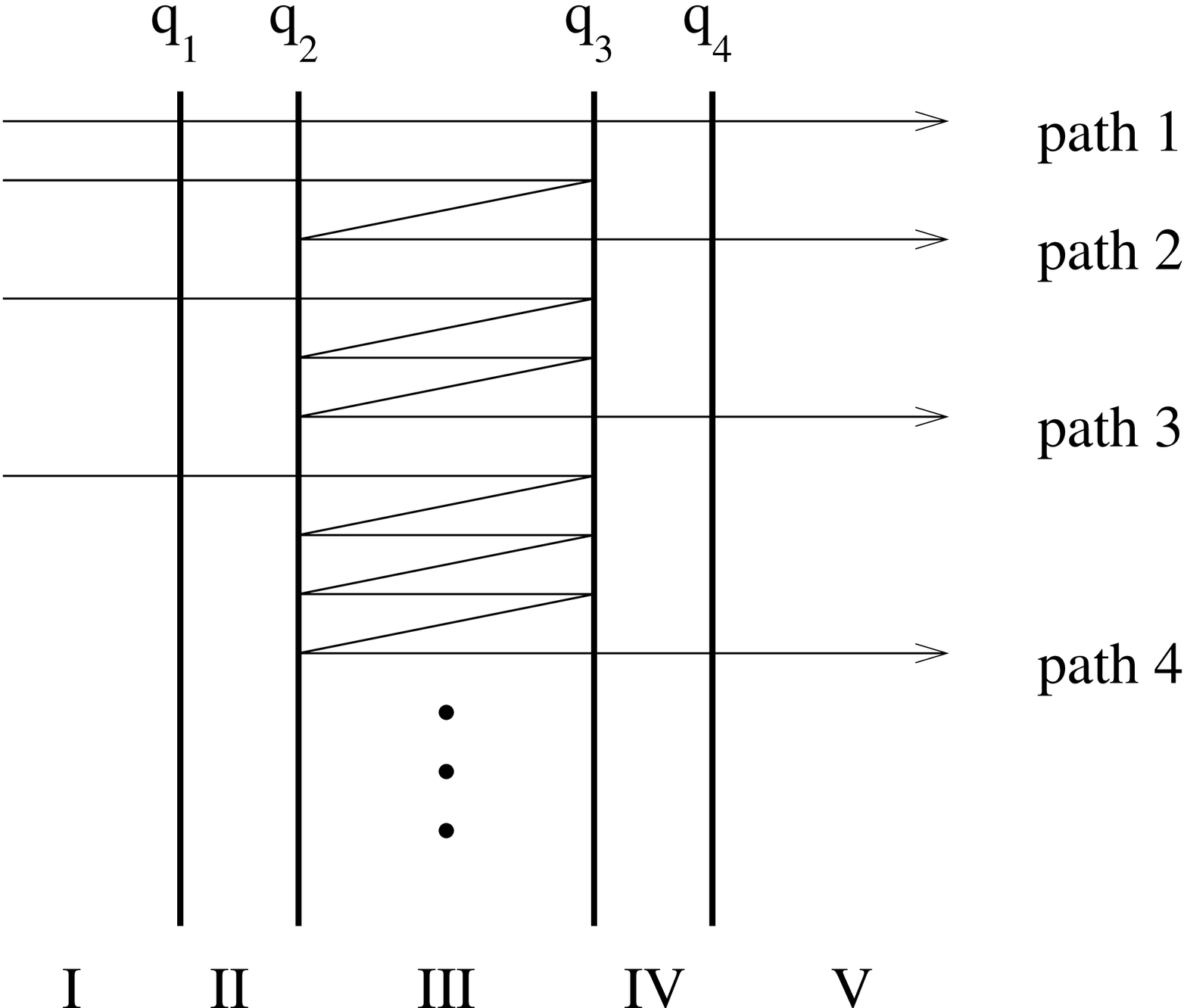}}
\caption{Some paths which contribute significantly to the path integral 
for the potential in Fig. \ref{fig:qmPot1}.} \label{fig:resPhase}}

Although there are other paths which contribute to the path integral, the ones in Fig. \ref{fig:resPhase}
are enough to demonstrate the phenomenon of resonant tunnelling. If the phase of these paths interfere constructively
upon leaving the system of barriers then the transmission amplitude becomes enhanced, and this can be understood
to be the condition that $2|q_2-q_3|$ corresponds to an integer number of de Broglie wavelengths. Within the WKB
approximation this requires
\ba
\label{eq:resTunCondQM}
W=\int_{q_2}^{q_3}p{\rm d}q=(n+\half)\pi,\;\;n\in\bZ,
\ea
where $p$ is the momentum. This condition is equivalent to saying that there exists a bound state in region III,
and corresponds to paths which oscillate back and forth in this region~\cite{Zhota:1990}.

Having understood the mechanism in quantum mechanics we need to apply this to field theory. This is achieved by
extending the N-dimensional quantum mechanics calculation \cite{Banks:1973ps} to field theory
\cite{Copeland:2007qf,Coleman:1977py,Bitar:1978vx,Sarangi:2007jb},
and has the effect that in the classically allowed regions the field obeys the classical equations, whereas in the
classically forbidden regions the field equations acquire a Wick-rotation to Euclidean time. As explained in~\cite{Copeland:2007qf}, we believe 
the field theory equivalent of the bound state is an oscillon, a localized field theory configuration which exhibits
periodic motion \cite{Gleiser:1993pt}. We are now in a position to understand the field theory version
of resonant tunnelling. To do this we consider a scalar theory with potential given by Fig. \ref{fig:qftPot1}, describing a theory with two local vacua ($A$ and $B$) and one global vacuum ($C$); note that the vacua $A$ and $C$
can be degenerate without affecting the argument.
\FIGURE{\centerline{\includegraphics[width=8cm]{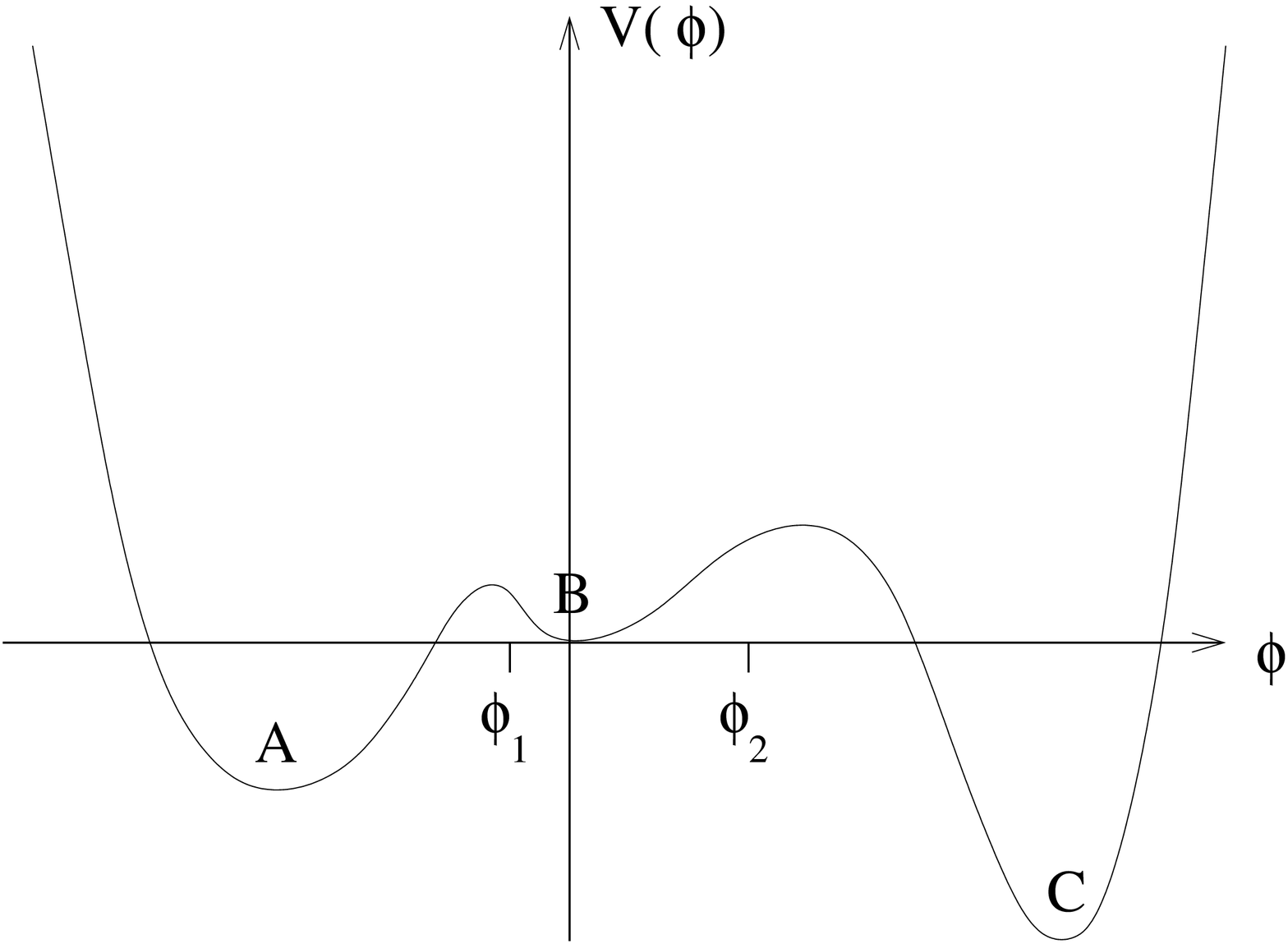}}
\caption{A potential with two false vacua, $\phi=\phi_A,\;\phi_B$, and one true vacuum, $\phi=\phi_C$.} \label{fig:qftPot1}}

In~\cite{Copeland:2007qf}, we considered tunnelling from a homogeneous region of one or other of the false vacua, $A$ or $B$, to the true vacuum, $C$. We now relax this assumption and consider the possibility that the initial state is a mixture of both false vacua. Physically we expect the field to asymptote towards the higher false vacuum, $B$, so that the initial state may be thought of as excited above the corresponding homogeneous solution. Now suppose that
there exists an oscillon solution\footnote{Oscillon profiles are typically rather close to a Gaussian shape.}
whose peak amplitude over an oscillation varies 
between $\phi_1$ and $\phi_2$. As vacuum $A$ has lower potential than $B$, then a bubble of $A$
vacuum surrounded by $B$ vacuum will expand, if it is of sufficient size. Similarly, a bubble of $C$ vacuum will also
expand if sufficiently large. This gives us all the ingredients we need to construct the field theory versions of the
paths in Fig. \ref{fig:resPhase}, so here we present a comparison between the particle and field theory cases.
\begin{itemize}
\item
Region I: In the mechanics case we set up the initial conditions of a particle moving toward the barriers. In field
theory we construct a bubble of $A$ vacuum and force it to contract.
\item
Region II: When the particle reaches the turning point we switch to Euclidean time and construct the MPEP from
$q_1$ to $q_2$. When the contracting bubble reaches its minimum size we switch to Euclidean time and evolve the
profile from this stationary state to the stationary oscillon state with amplitude $\phi_1$.
\item
Region III: At turning point $q_2$ we return to normal time and allow the particle to oscillate back and
forth between $q_2$ and $q_3$. The particle can oscillate any number of times, with each possibility giving
a contribution to the path integral as in Fig. \ref{fig:resPhase}. For the field theory, we go back to a Lorentzian
signature and allow the oscillon to evolve. Again, it may perform any number of oscillations, with each case contributing
to the path integral.
\item
Region IV: After the particle has oscillated in region III we revert to Euclidean time and evolve it between $q_3$ and
$q_4$. For the field theory we resume a Euclidean signature and evolve from the oscillon profile with amplitude $\phi_2$
to a bubble of $C$ vacuum surrounded by $B$ vacuum.
\item
Region V: The final part of the motion sees the equations return to normal time, and the particle rolls from
$q_4$ to $+\infty$. While for the field, we again switch back to Lorentzian signature and watch the bubble of $C$
vacuum expand and consume the surrounding $B$ vacuum. 
\end{itemize}

There are a number of difficulties in achieving this within an actual model, not least of which 
is the rather special initial state of a contracting spherical bubble which would prefer to expand. 
There is then the task of finding a field theory with periodic oscillon solutions. Although oscillons have been found to
be long-lived compared to natural timescales \cite{Gleiser:1993pt} they are typically not
strictly periodic. One can construct periodic solutions using a Fourier expansion \cite{Saffin:2006yk,Honda:2001xg} but
these tend to contain incoming radiation from infinity in order to counter-balance the radiation being emitted. 
Once we have a model with a periodic oscillon we need to make sure that the Euclidean evolutions (regions II and IV)
join the peak-amplitude-profiles onto the (momentarily) static bubbles of vacuum $A$ and $C$ - a far from trivial requirement.

After these conditions have been met we can then ask what will make tunnelling from the $A$ bubble to the $C$
bubble resonate. One finds that in d spatial dimensions, (\ref{eq:resTunCondQM}) gets generalized to \cite{Copeland:2007qf,Sarangi:2007jb}
\ba
\label{eq:resTunCondQFT}
W=\int_{t_1}^{t_2}{\rm d^d}x{\rm d}t\;\dot\phi^2=(n+\half)\pi,\;\;n\in\bZ,
\ea
where $t_1$ and $t_2$ are the times for the half-oscillation of the oscillon. To answer the question what is the probablity
of tunnelling from bubble $A$ to bubble $C$? we need to introduce the Euclidean actions for regions II and IV.
\ba
\label{eq:EucAct}
\sigma_{II}=\int_{II}{\rm d^d}x{\rm d}\tau\;(\phi_\tau)^2,\\
\sigma_{IV}=\int_{IV}{\rm d^d}x{\rm d}\tau\;(\phi_\tau)^2,
\ea
where $\phi_\tau=\frac{d\phi}{d\tau}$, and $\tau$ is the Euclidean time parameter in the equations of motion for the MPEP.
The transition probability is then given by \cite{Merz,Tye:2006tg,Copeland:2007qf,Sarangi:2007jb}
\ba
\label{eq:tranProb}
T_{I \to V}=4\left\{\left[\Theta_{II}\Theta_{IV}+1/(\Theta_{II}\Theta_{IV})\right]^2\cos^2W
         +\left[\Theta_{II}/\Theta_{IV}+\Theta_{IV}/\Theta_{II}\right]^2\sin^2W\right\}^{-1},
\ea
where we define $\Theta_{II}=\exp(-\sigma_{II})$, $\Theta_{IV}=\exp(-\sigma_{IV})$. From this expression we see that
$T_{I \to V}$ has peaks when (\ref{eq:resTunCondQFT}) is satisfied.

\section{Finding a model} 
\lab{model}
The first thing we shall look for is a model which contains strictly periodic oscillons. In one
spatial dimension there exists breather solutions of the sine-Gordon model which are the archetype of oscillons.
Going to higher dimensions one finds that strictly periodic solutions are rather harder to find 
\cite{Saffin:2006yk,Gleiser:2004an}, with solutions being long lived but typically only quasi-periodic
owing to them radiating. 
However, there is a model which does have strictly periodic solutions in general spatial dimensions. 
This theory has already proved a useful testing ground for a number of non-perturbative phenomena such as the
study non-linear waves, zero temperature tunnelling, finite temperature tunnelling, 
\cite{bbm}
and is given by the following Lagrangian,
\ba
\hat{\cal L}&=&-\half\hat\del_\mu\hat\phi\hat\del^\mu\hat\phi
               -\half m^2\hat\phi^2\left[1-\ln\left(\hat\phi^2/c^2\right)\right].
\ea
This can be used to describe a system similar to Fig. \ref{fig:qftPot1} where the vacua $A$ and $C$ are very deep.
We shall work with the dimensionless variables $\phi=\hat\phi/c$, $x^\mu=m\hat x^\mu$, such that the equation
of motion in $d$ spatial dimensions, for a radial profile is given by
\ba
\ddot\phi-\phi''-\frac{d-1}{r}\phi'=\phi\ln\phi^2,
\ea
where $\dot\phi \equiv \frac{d\phi}{dt}$ and $\phi' \equiv \frac{d\phi}{dr}$. 
The remarkable property of this equation is that, despite being non-linear, it is separable with the normalizable
solution given by
\ba
\phi(t,r)&=&T(t)\exp(-r^2/2)
\ea
where $T$ satisfies the following o.d.e. in the classically allowed region
\ba
\ddot{T}&=&-\frac{d}{dT}\left[\half(1+d)T^2-\half T^2\ln T^2\right]=-\frac{d}{dT}V_L(T),
\ea
and along the MPEPs the function $T$ satisfies
\ba
\frac{d^2T}{d\tau^2}&=&-\frac{d}{dT}\left[-\half(1+d)T^2+\half T^2\ln T^2\right]=-\frac{d}{dT}V_E(T).
\ea
The $L$ and $E$ subscripts refer to Lorentzian and Euclidean respectively. While these equations for $T$
have not proved tractable analytically, they are simple to solve numerically and constitute a significant
simplification of the field theory.
\FIGURE{\centerline{\includegraphics[width=12cm]{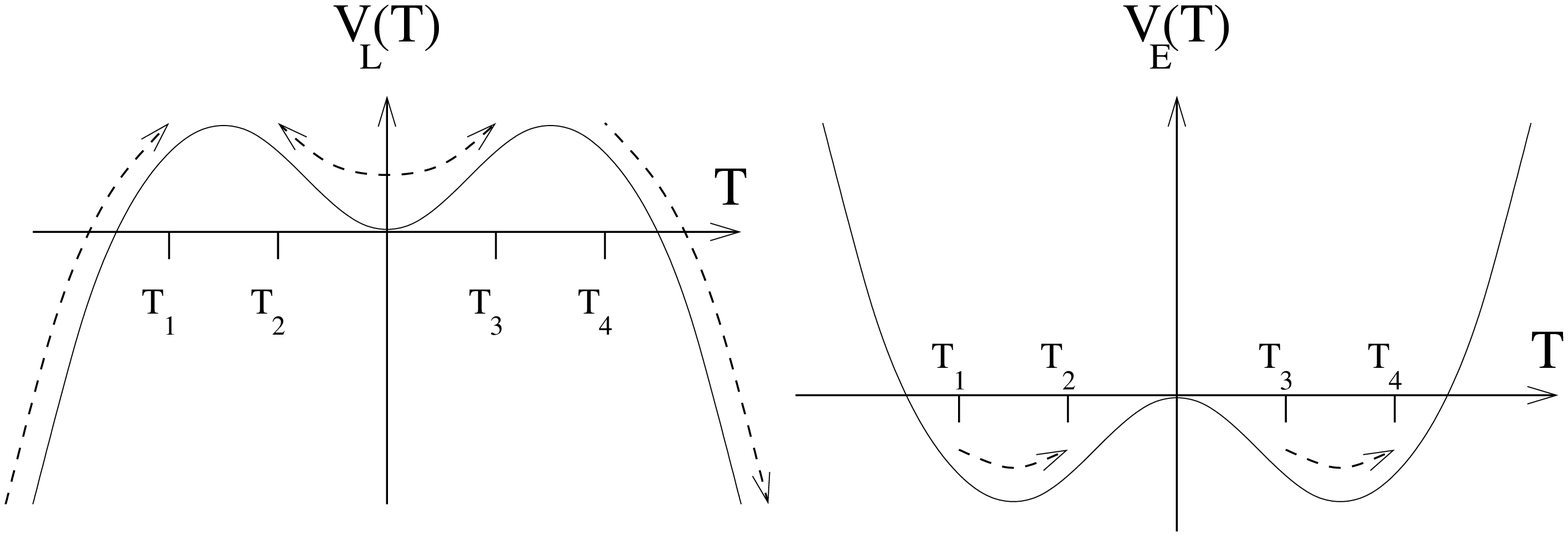}}
\caption{The potentials determining the evolution of $T$. The Lorentzian potential, $V_L$ applies in the classically allowed regions, whereas the Euclidean potential, $V_E$, applies in the classically forbidden regions.} \label{fig:V_EV_L}}

As the system has now been separated we can consider it in terms of the mechanics of a particle with position $T$, 
moving in a potential $V_L$ for the classically allowed region, and $V_E$ along a MPEP.  The full evolution 
for both the classical solution and the tunnelling solution can be understood by considering Fig. \ref{fig:V_EV_L}. The particle starts on the Lorentzian potential, $V_L$, at large negative $T$ and evolves up to the point $T_1$. Classically, it will turn around at this point and start to roll back down $V_L$. This gives rise to the classical trajectory shown in Fig. \ref{fig:qftClassical}.
\FIGURE{\centerline{\includegraphics[height=3.2cm, width=4.5cm]{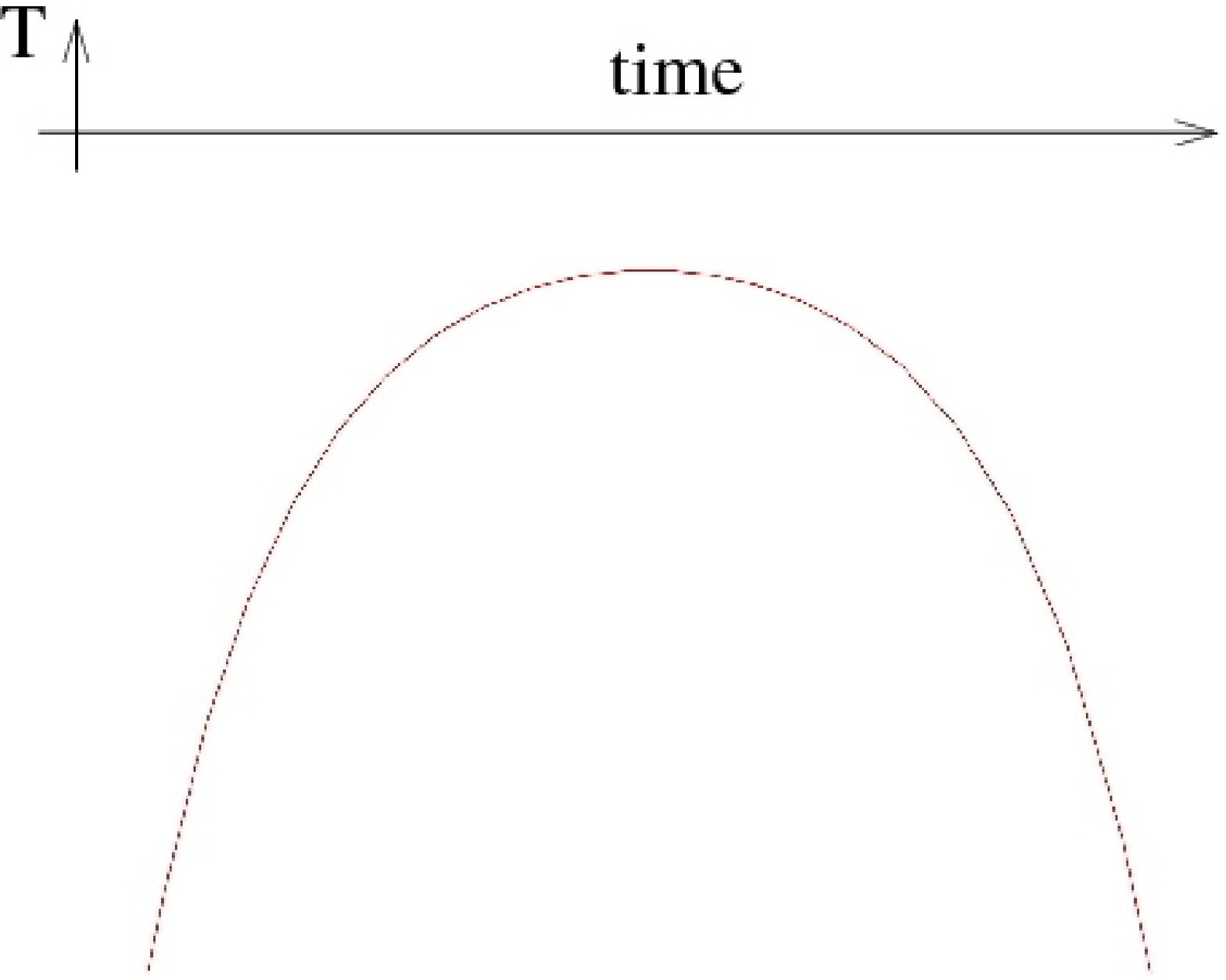}\includegraphics[height=3.2cm, width=6.4cm]{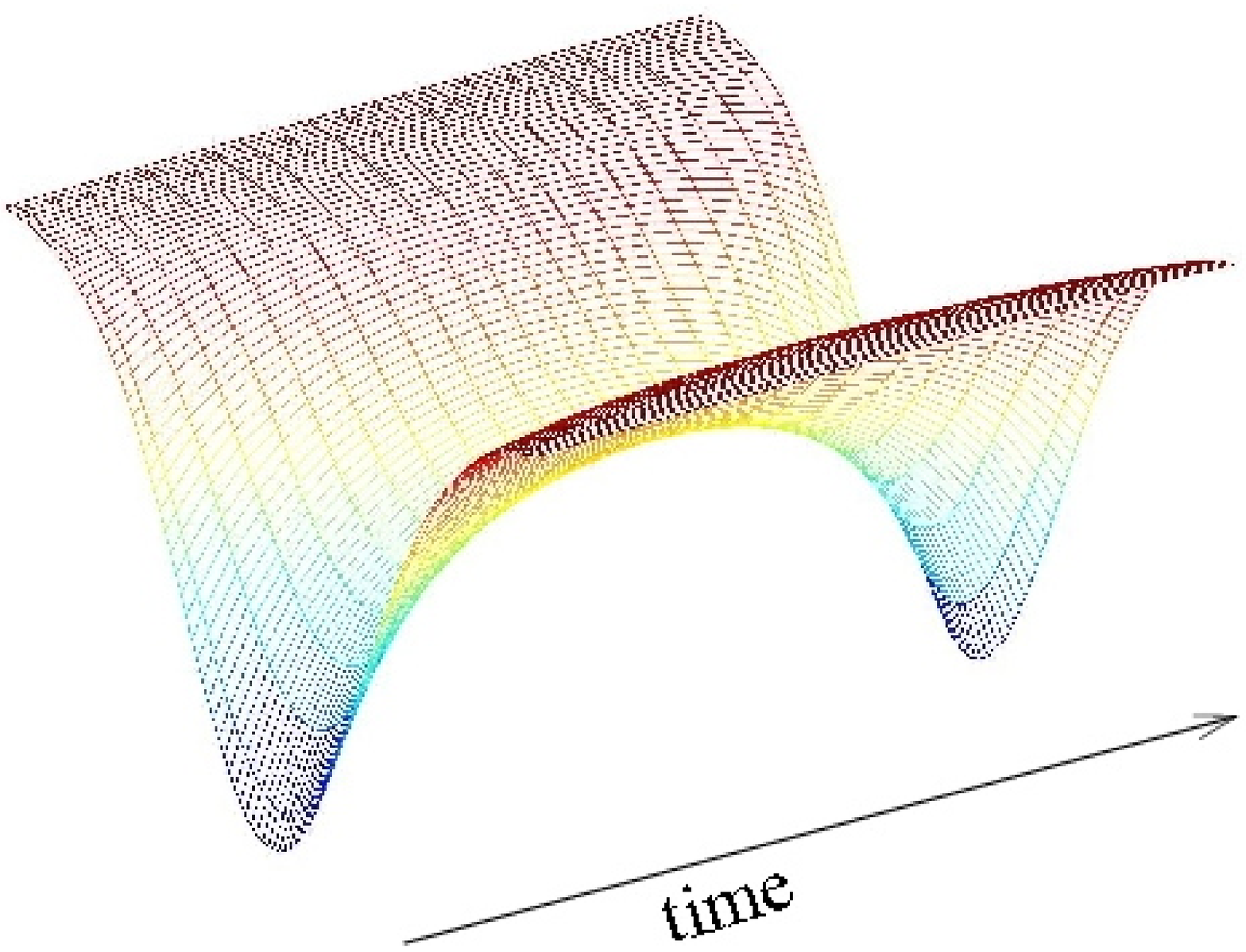}}
\caption{The classical evolution of an initially contracting bubble. The evolution parameter is, of course, real time, $t$. The lefthand column shows $T(t)$ whereas the righthand column shows $\phi(t, r)$. Classically, the bubble shrinks to its minimum size before expanding out again.} \label{fig:qftClassical}}
In contrast, the tunnelling trajectories behave rather differently. At $T_1$, the particle tunnels along a MPEP towards the next turning point at $T_2$, under the influence of the Euclidean potential, $V_E$. At $T_2$, the particle 
returns to the classically allowed region and the influence of the Lorentzian potential. At this stage it oscillates some number of times between $T_2$ and $T_3$, before tunnelling again along another MPEP between $T_3$ and $T_4$. Finally, classical evolution takes over at $T_4$ and $T$ continues to roll
down $V_L$. 
\FIGURE{\leftline{\small Path 1:}
\centerline{\includegraphics[height=3.2cm, width=4.5cm]{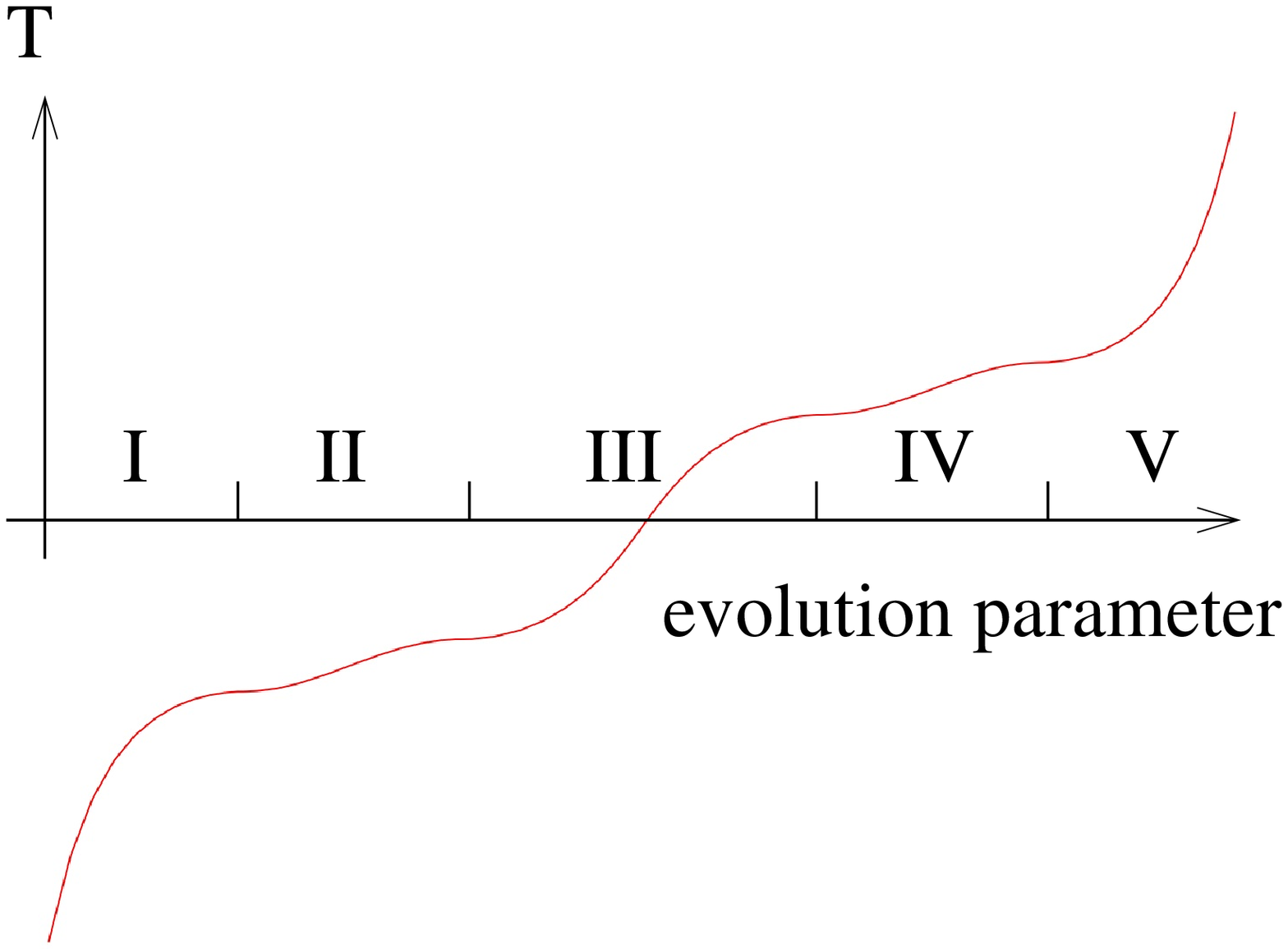}\includegraphics[height=3.2cm,width=6.4cm]{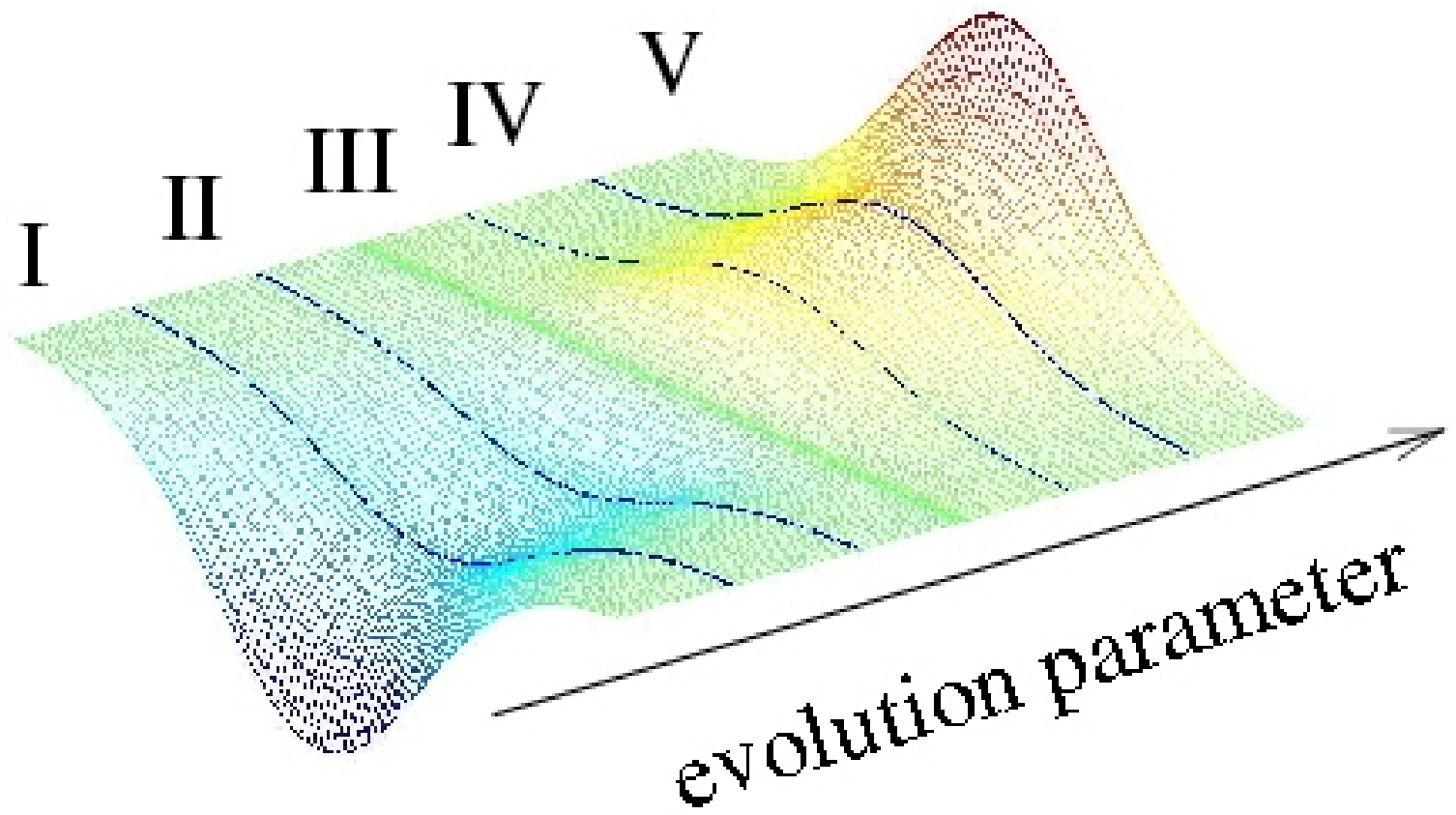}}
\leftline{\small Path 2:}
\centerline{\includegraphics[height=3.2cm,width=4.5cm]{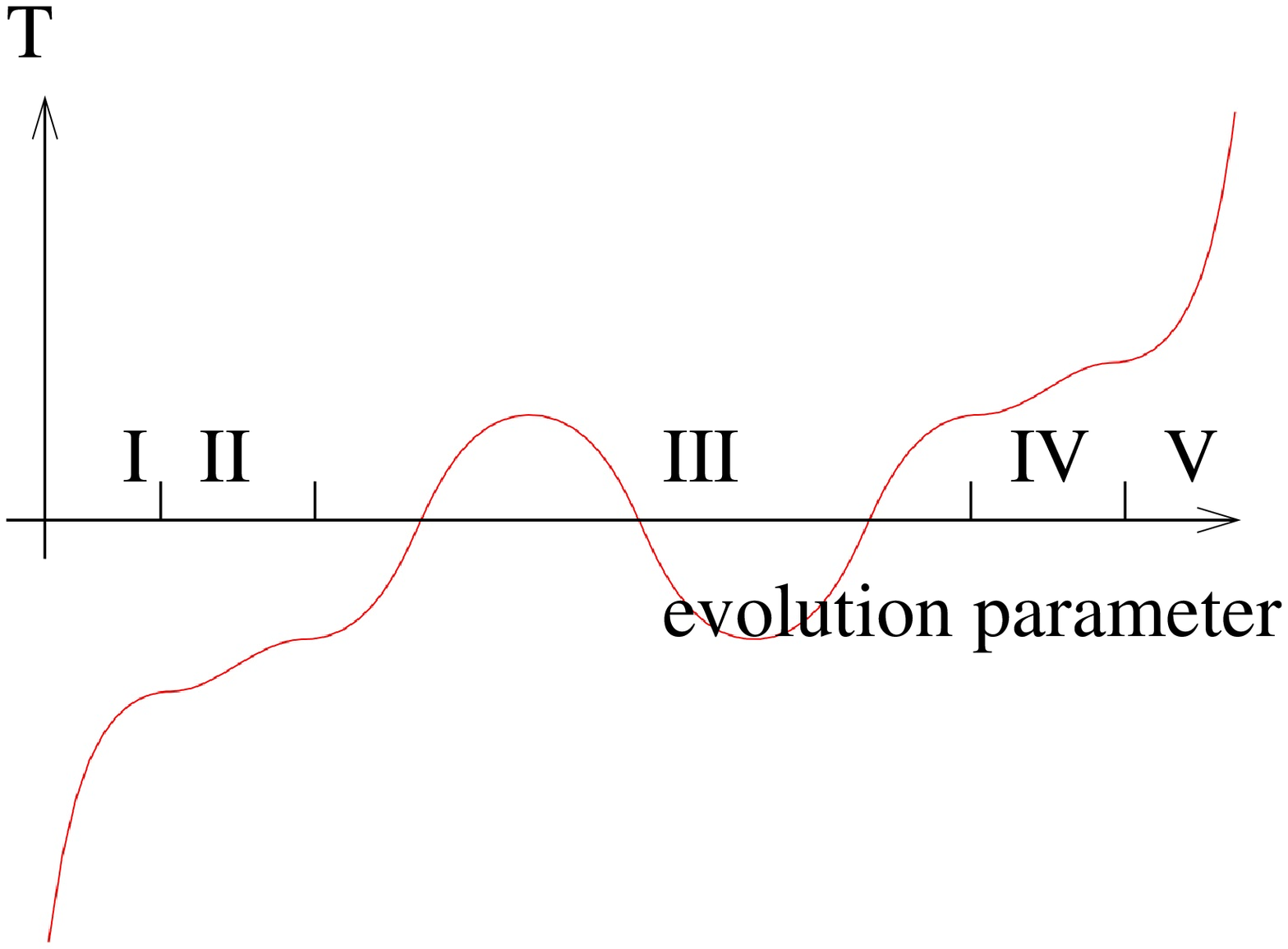}\includegraphics[height=3cm,width=6.4cm]{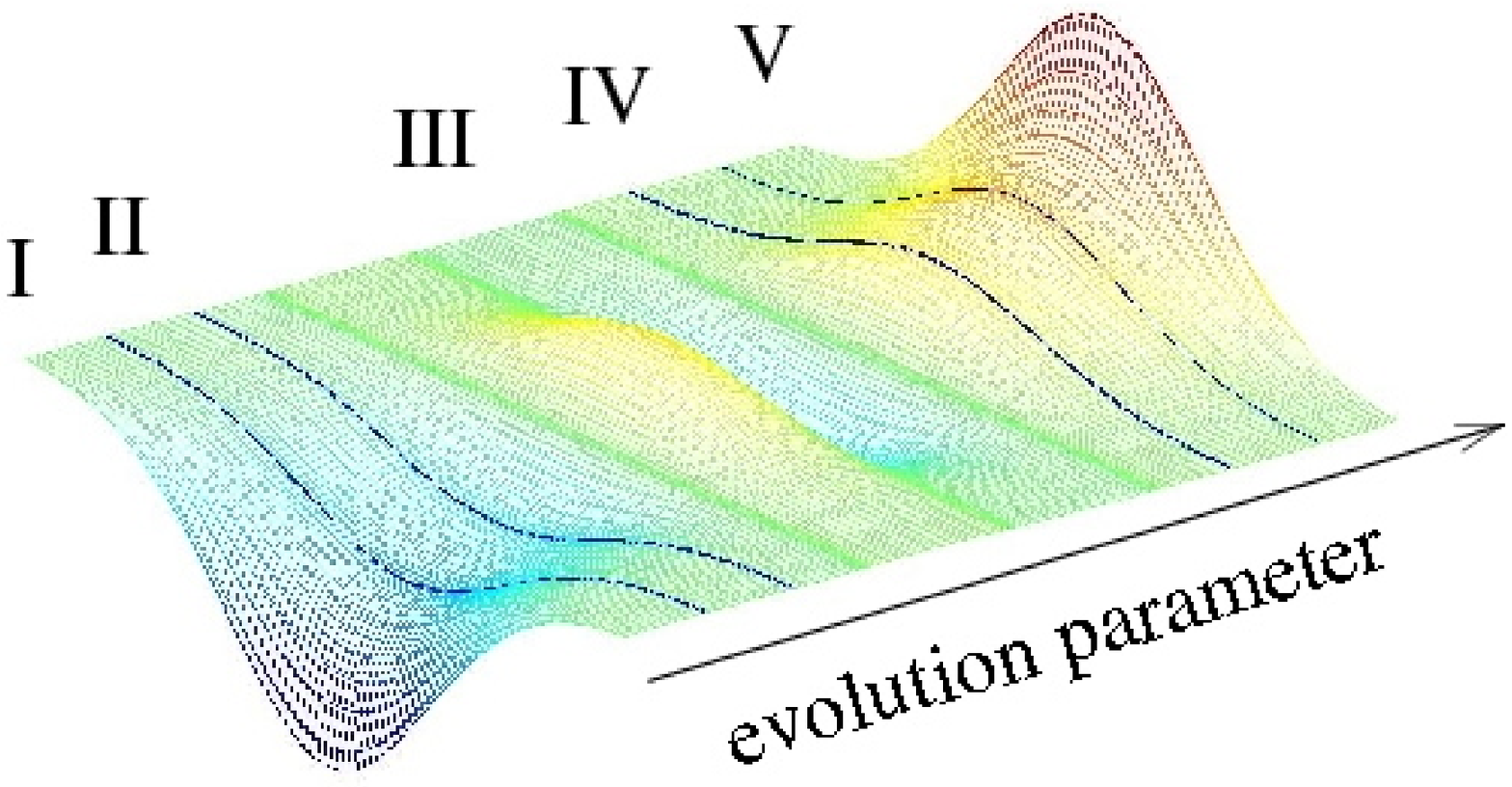}} 
\leftline{\small Path 3:}
\centerline{\includegraphics[height=3.2cm,width=4.5cm]{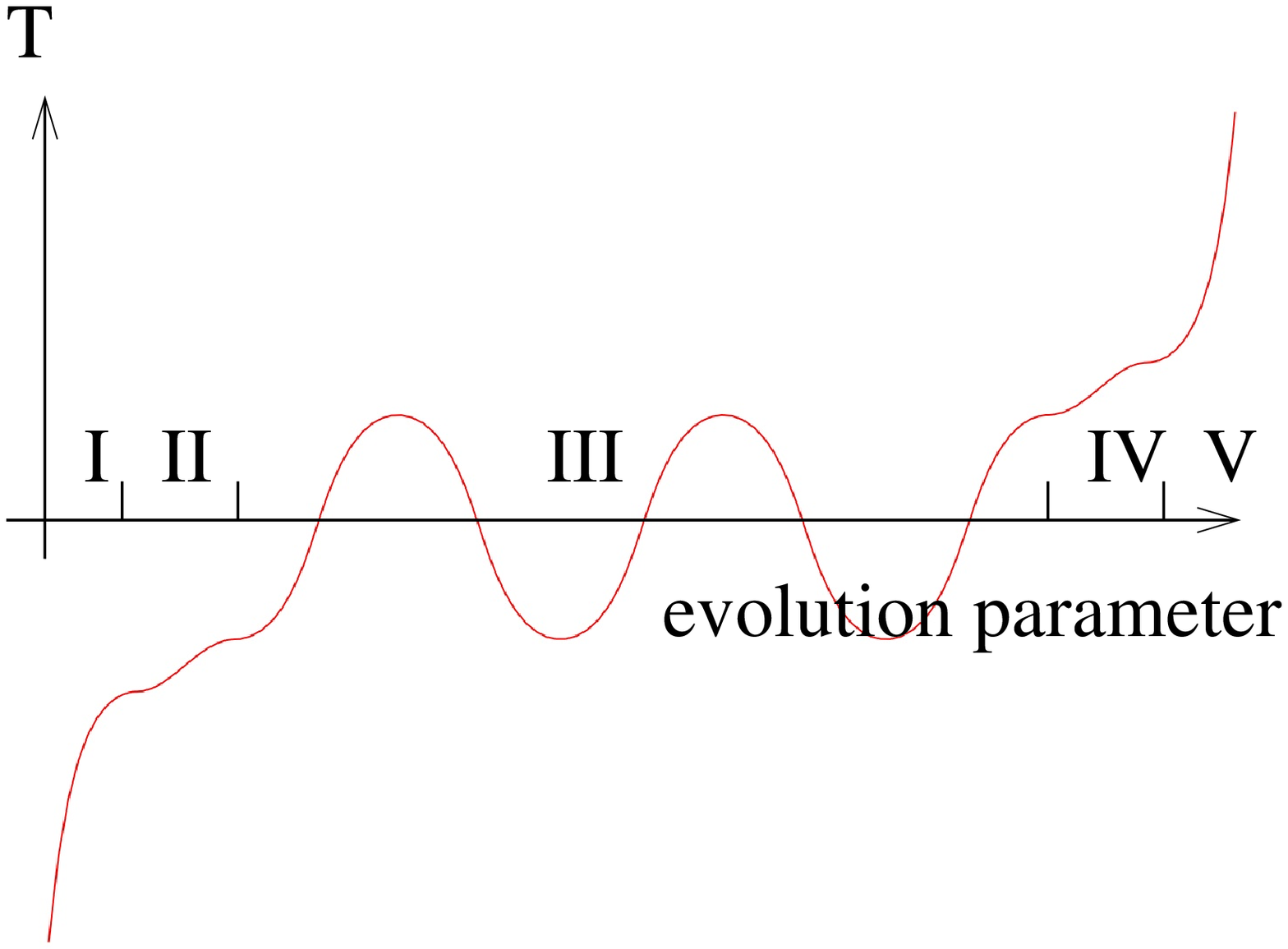}\includegraphics[height=3.2cm,width=6.4cm]{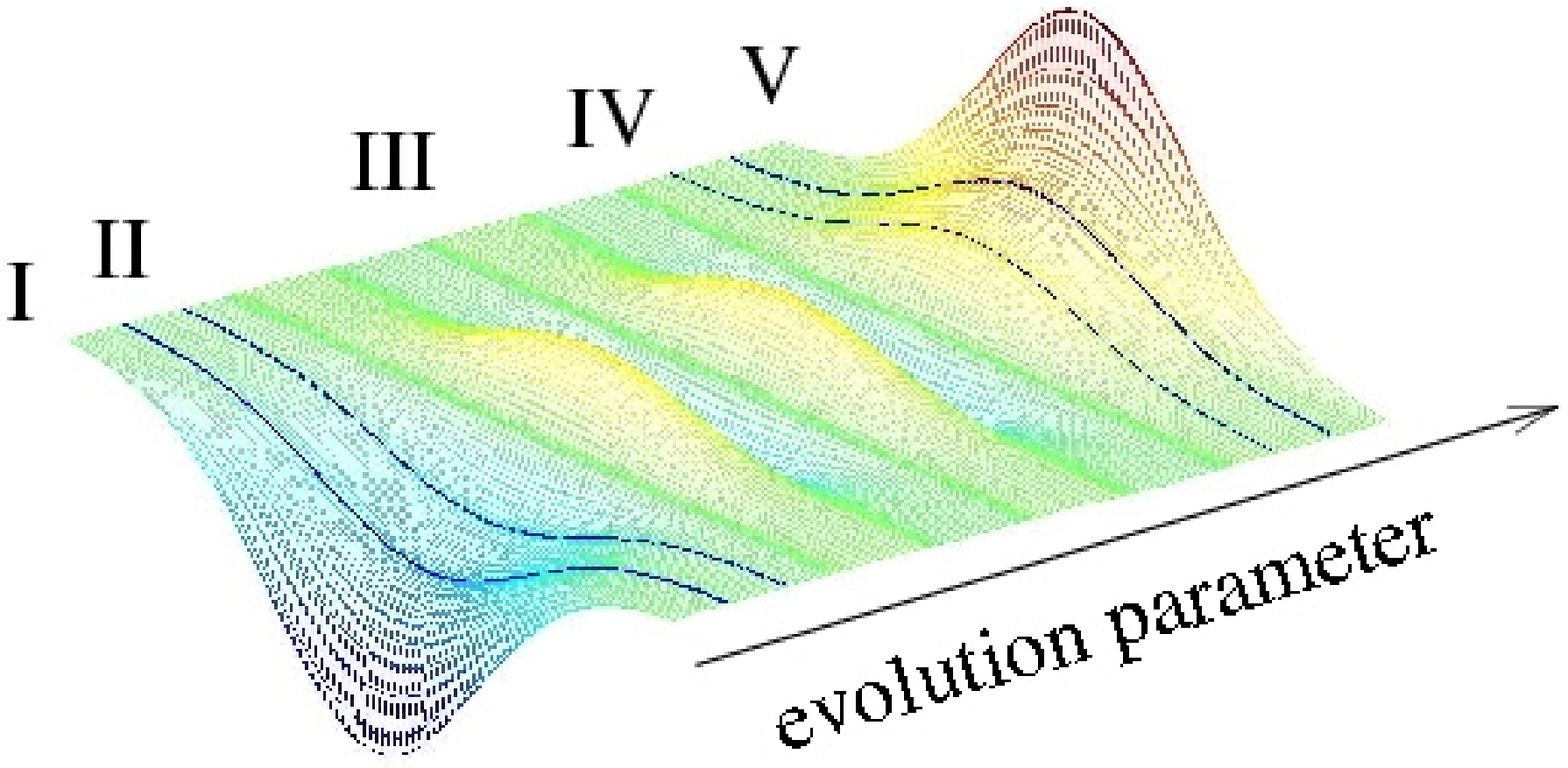}}
\leftline{\small Path 4:}
\centerline{\includegraphics[height=3.2cm,width=4.5cm]{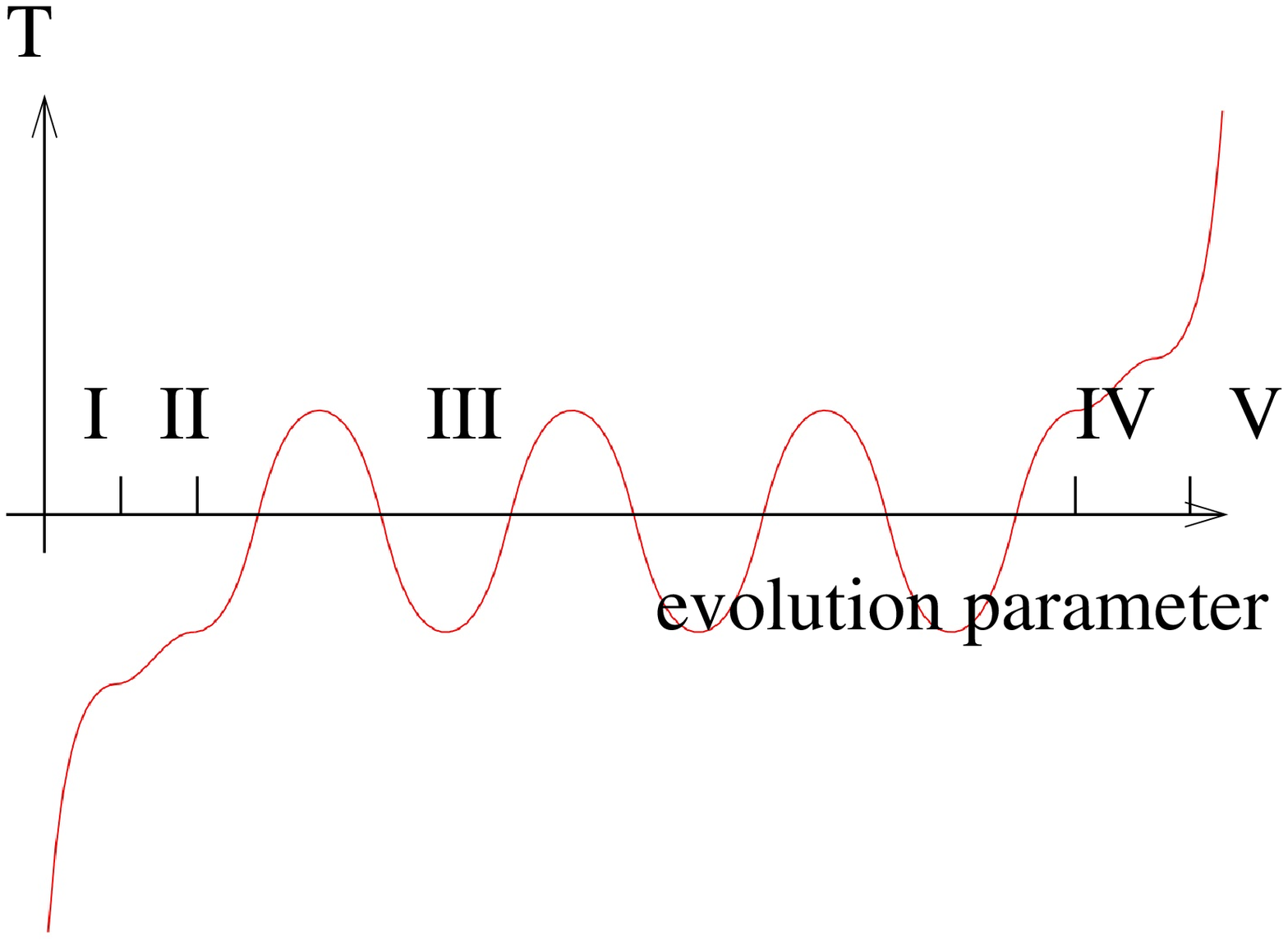}\includegraphics[height=3.2cm,width=6.4cm]{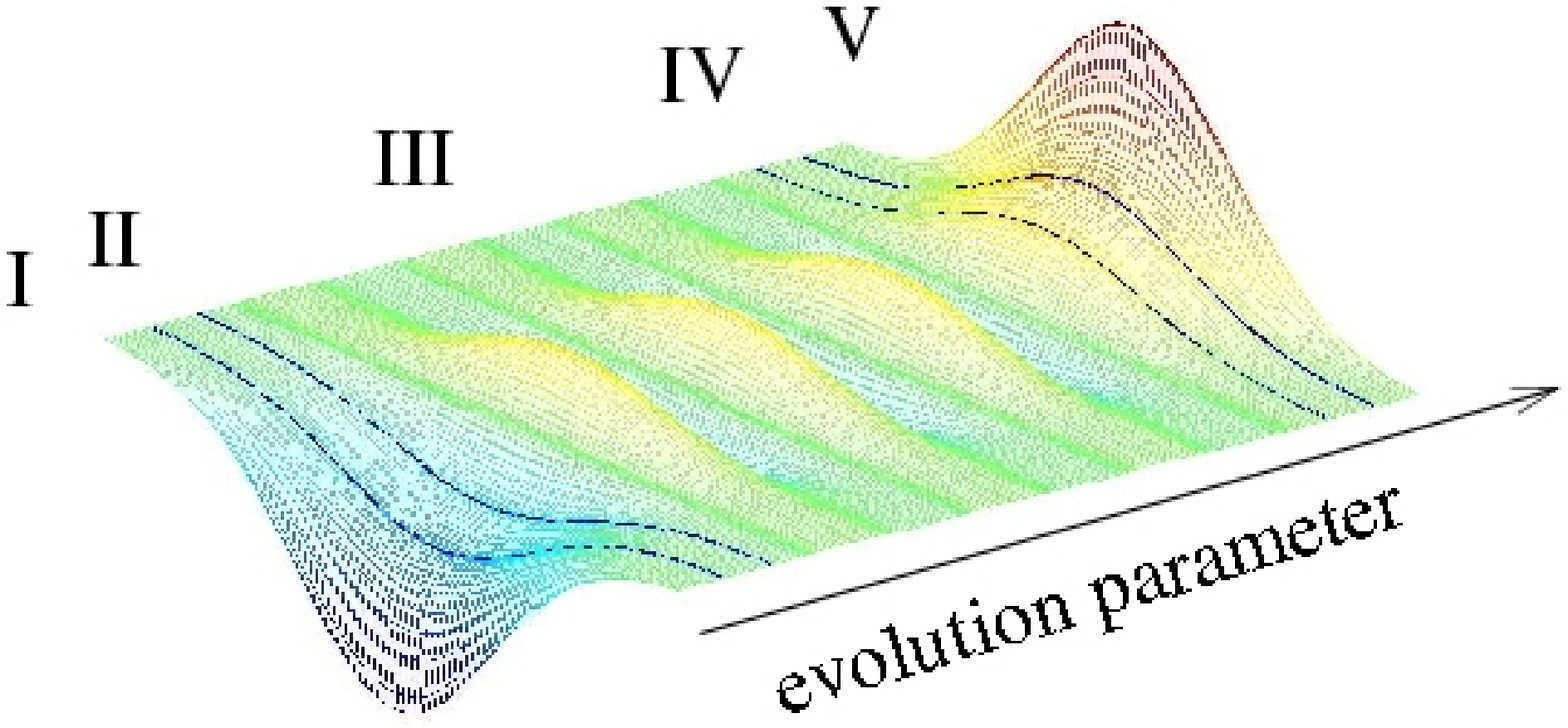}}
\caption{Some configurations, or ``paths'', which contribute to the path integral for field theory resonant tunnelling. The lefthand column shows the evolution of $T(\la)$ along a particular path, whereas the righthand column shows the evolution of the entire field, $\phi=T(\la) e^{-r^2/2}$. The evolution parameter, $\la$, is given by real time, $t$ in the classically allowed regions, I, III and V, and by Euclidean time, $\tau$ in the classical forbidden regions, II and IV. The ``paths'' are labelled $1$ to $4$ in reference to their quantum mechanical counterparts (see Fig. \ref{fig:resPhase}).} \label{fig:qftPathInt}}

These five different regions of evolution are the precise analogue of the regions used in quantum
mechanics to calculate resonant tunnelling (see Fig. \ref{fig:resPhase}). Each of the particle paths in Fig. \ref{fig:resPhase}
get translated into the field theory evolutions of Fig. \ref{fig:qftPathInt}. For example, path 1 of Fig. \ref{fig:qftPathInt}
shows a solution of the field theory which starts with a contracting Gaussian bubble,
evolving in Lorentzian time (region I), which reaches a minimum size whereupon we switch to Euclidean time and evolve the
field in region II to the next turning point. From there we return to Lorentzian time (region III)
and perform half an oscillation of the oscillon and match to region IV for the next stage of Euclidean evolution
that finally joins to region V for the Lorentian evolution of an expanding bubble. So evolution 1 of 
Fig. \ref{fig:qftPathInt} is the field theory version of path 1 in Fig. \ref{fig:resPhase}. Similarly, evolutions
2, 3, 4 of Fig. \ref{fig:qftPathInt} are the field theory equivalents of paths 2, 3, 4 in Fig. \ref{fig:resPhase},
each with a different number of oscillations in region III. These compare to the classical evolution for the same initial conditions
which is presented in Fig. \ref{fig:qftClassical}. For the classical evolution we start with a 
contracting bubble with Gaussian profile and negative amplitude which evolves to an expanding bubble with Gaussian
profile and negative amplitude. In contrast, the tunnelling trajectories 1 to 5 each take us from a negative amplitude to a positive amplitude
Gaussian bubble.
\section{The transmission rate} 
\lab{transRate}
To calculate the transmission rate for the double barrier system in quantum mechanics, using the WKB approximation, we need the
following integrals evaluated between the stationary turning points separating classically allowed and forbidden regions
\ba
\sigma_{II}&=&\int_{\tau_1}^{\tau_2}{\rm d}^dx{\rm d}\tau\;\left(\frac{d}{d\tau}\phi\right)^2,\\
W&=&\int_{t_2}^{t_3}{\rm d}^dx{\rm d}t\;\dot\phi^2,\\
\sigma_{IV}&=&\int_{\tau_3}^{\tau_4}{\rm d}^dx{\rm d}\tau\;\left(\frac{d}{d\tau}\phi\right)^2.
\ea
In regions II and IV, the MPEPs are parametrised using Euclidean time, $\tau$, whereas in region III the classical path is parametrised using real time $t$. These integrals may be thought of as functions of the energy, as we will explain shortly. Energy remains constant throughout the motion, and is defined relative to the homogeneous vacuum, $\phi \equiv 0$. In the initial, classically allowed, region, it is given by 
\be
E=m^{d+2} c^2 \int d^d x ~\half \dot \phi^2+\half (\vec \nabla \phi)^2+\half \phi^2(1-\ln\phi^2) =\half m^{d+2} c^2\Gamma\left(\frac{d}{2}\right)\Omega_{d-1}\left[ \half\dot T^2+V_L(T)\right]
\ee
where $\Omega_{d-1}$ is the volume of the unit $(d-1)$-sphere. The energy is strictly positive for excited initial states, a property which distinguishes the decay of inhomogeneous states discussed here from the homogeneous false vacuum decay discussed in~\cite{Copeland:2007qf}.

To see how $\sigma_{II},~W$ and $\sigma_{IV}$ depend on $E$, recall that the classical turning point occurs when
$E=U[\phi_{tp}]$, where \cite{Copeland:2007qf}
\be \lab{tp}
U[\phi]=m^{d+2} c^2\int d^dx ~\left[\half (\vec \nabla \phi)^2+\half \phi^2(1-\ln\phi^2)\right]
\ee
Given that $E$ is fixed throughout the motion, to find the turning points in any given region, II, III, or IV, we simply plug the relevant field configuration into equation (\ref{tp}) and solve for real or Euclidean time as appropriate. In  region II, for example, this gives $\tau_1=\tau_1(E), ~\tau_2=\tau_2(E)$, and so $\sigma_{II}=\sigma_{II}(E)$.

Given the symmetry of the potential, $V_E$, we immediately see that $\sigma_{II}=\sigma_{IV}$.  Therefore, by defining
$\Theta=\exp(\sigma_{II})$ one has that the transmission probability is given by\ba
\label{eq:BBMtrans}
T_{I \to V}=4\left[\left(\Theta^2+\Theta^{-2}\right)^2\cos^2W+4\sin^2W\right]^{-1}
\ea
It is easy to see that the transmission rate is enhanced as $W(E) \to (n+\half)\pi$, for any integer, $n$. The width of this resonance can be approximated by \cite{Merz}
\ba
\label{eq:resWidth}
\Gamma&\simeq&\left(\pi\Theta^2\frac{dW}{dE}\right)^{-1}=2\left(\pi\Theta^2\Delta t\right)^{-1}.
\ea
where we have used the fact from classical mechanics that the period of oscillation
is given by\footnote{The factor of two is because we define $W$ over a half period.}
$\Delta t=2\frac{dW}{dE}$.

To complete the analysis of this section we present the results of this model, taking $d=3$ spatial dimensions, and
using $m=c=1$. Fig. \ref{fig:functions} shows how the quantities $W$, $\sigma_{II}$ and $\cos(W)$  depend
on energy, from which we may calculate the transmission probability. In Fig. \ref{fig:transmission} we give
a section of the curve showing the transmission probability,  noting the presence of resonance peaks which become
increasingly narrow as energy is reduced. As we go to energies lower than those shown in Fig. \ref{fig:transmission}, this behaviour continues, with increasingly narrow  resonant peaks appearing at regular intervals.  From equation (\ref{eq:resWidth}), we might have expected the width of the resonance to decrease with energy \ie as $\sigma_{II}$ increases. Intuitively this also makes sense: $\sigma_{II}$ essentially measures the height of the potential barrier, and the higher the barrier the more tunnelling is suppressed, except in the very core of the resonance.

Fig. \ref{fig:transmission} actually shows the transmission rate in the region immediately  below  a maxmium energy, $E_{max}=m^{d+2} c^2\pi^{d/2}e^d/2$ ($\sim 56$ for $d=3, ~m=c=1$), beyond which our tunnelling description breaks down. This is because at high enough energies the barrier disappears altogether and one can pass from one vacuum to the other classically. At lower energies, of course, one must tunnel between vacua quantum mechanically, with transmission probability given by (\ref{eq:BBMtrans}). Although this process is usually suppressed, the resonant peaks in Fig. \ref{fig:transmission} demonstrate that an inhomogeneous initial state {\it can} decay via resonant tunnelling, passing from one vacuum to another with almost unit probability. Note that for $E=0$, there is no resonance, which is consistent with our original no-go theorem~\cite{Copeland:2007qf}. We conclude, therefore, that initial inhomogeneity and non-zero energy are absolutely crucial for resonant tunnelling to occur in quantum field theory.

\FIGURE{\centerline{\includegraphics[width=8cm]{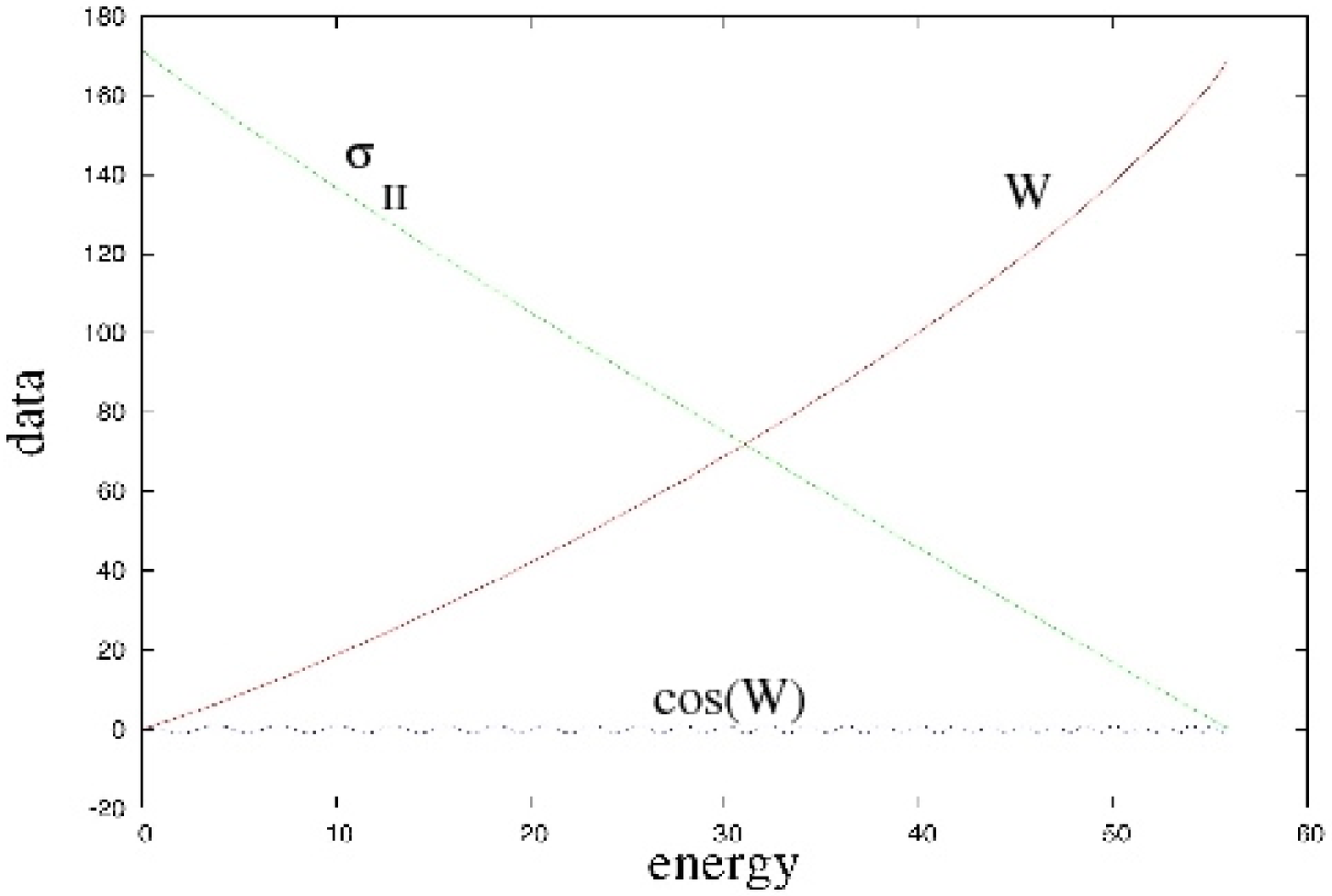}}
\caption{A plot of the quantities used to calculate the transmission probability (\ref{eq:BBMtrans}).} \label{fig:functions}}
\FIGURE{\centerline{\includegraphics[width=8cm,angle=-90]{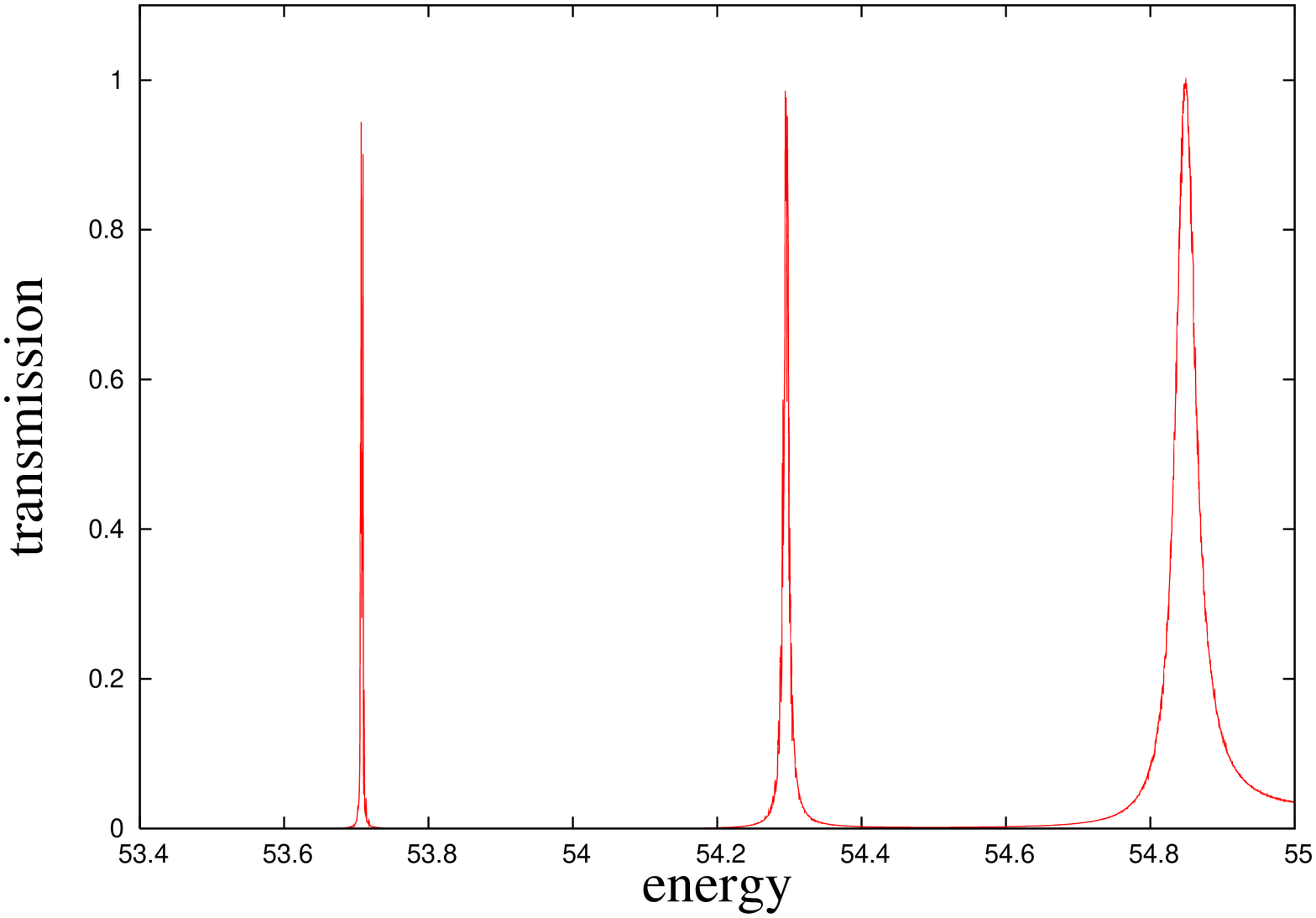}}
\caption{A graph showing the dependence of transmission probability on energy for a range of energies, just below the maximum energy, $E_{max} \sim 56$. Notice the presence of the resonant peaks at regular intervals, the width of which decreases with energy.} \label{fig:transmission}}

\section{Thin wall limit} 
\lab{sec:thinWall}
How typical are the results of the previous section?  This is not so easy to answer in complete generality, although we can make some progress by restricting attention to theories for which there is a sensible ``thin wall limit'', as is often assumed in tunnelling calculations~\cite{Coleman:1977py}. Although the  previous calculation made use of a model containing oscillons with a single length scale, namely its size, in principle it is possible to have oscillons whose radius is much larger than the width of their ``skin''. In this instance, the thin wall approximation is a good one, and the dynamics of the oscillon is
governed by a membrane-type action
\ba \lab{thinaction}
S_m&=&-\sigma\int{\rm d}^3\xi\sqrt{{\rm det}G}+\epsilon\int{\rm d}V{\rm d}t.
\ea
where $G$ is the induced metric on the world volume of the membrane, $\sigma$ the tension of the bubble wall,
$V$ the volume of the bubble and $\epsilon$ the difference in potential energy density between the inside and outside of the bubble. Even without our general motivation, we could just take (\ref{thinaction}) as our starting point and consider the dynamics of membranes
in their own right. Imposing spherical symmetry,  the action (\ref{thinaction})  of a single bubble wall leads to the following Lagrangian for the
 bubble radius, 
\ba
L_{tw}&=&-\Omega_{d-1}\sigma R^2\sqrt{1-\dot{R}^2}+\Omega_d\epsilon|R|^3.
\ea
We can extract the potential~\cite{Copeland:2007qf},
\ba
V_{tw}&=&\Omega_{d-1}\sigma R^2-\Omega_d\epsilon|R|^3
\ea
which is qualitatively the same as $V_L(T)$ of the scalar field theory. This  means that
we can simply import qualitative results from the previous section.

In the thin wall 
analysis of our previous paper~\cite{Copeland:2007qf}, we pointed out that for resonance to occur, we need an oscillatory solution to act as the intermediate bound state. In the case of tunnelling
from the false vacuum ($R=0$) we see that there are no such solutions, as expected from our no-go theorem~\cite{Copeland:2007qf}.  However, if we allow for {\it in}homogeneous initial conditions  we may consider a range of energies for which $V_{tw}$ supports oscillatory
solutions. Indeed, one can describe a process whereby a contracting bubble of one vacuum decays, via resonant tunnelling, to an expanding bubble of a completely different vacuum. The oscillatory intermediate state acts as a springboard for the tunnelling amplitude. 
\FIGURE{\centerline{\includegraphics[height=4cm, width=6cm]{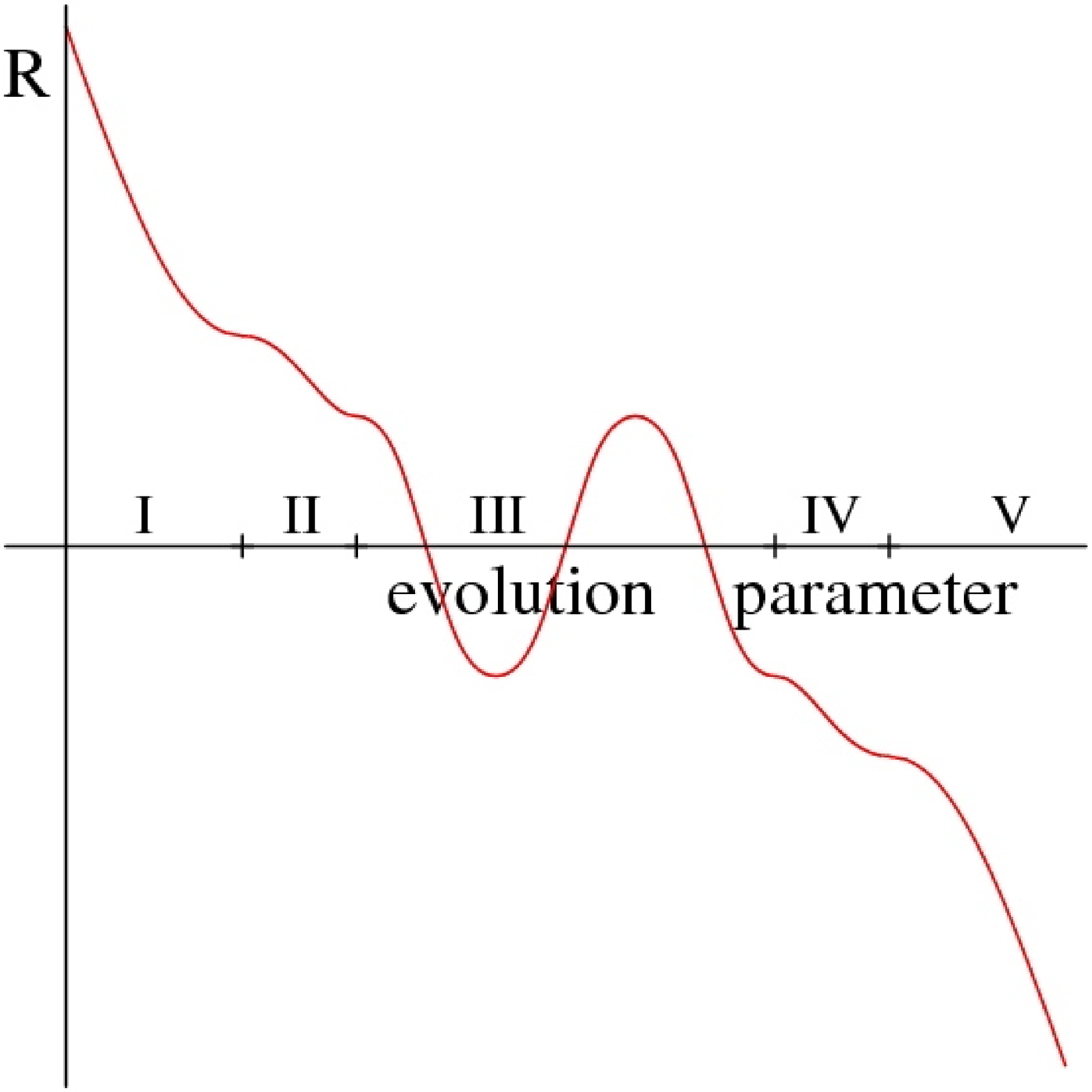}\includegraphics[height=4cm, width=8cm]{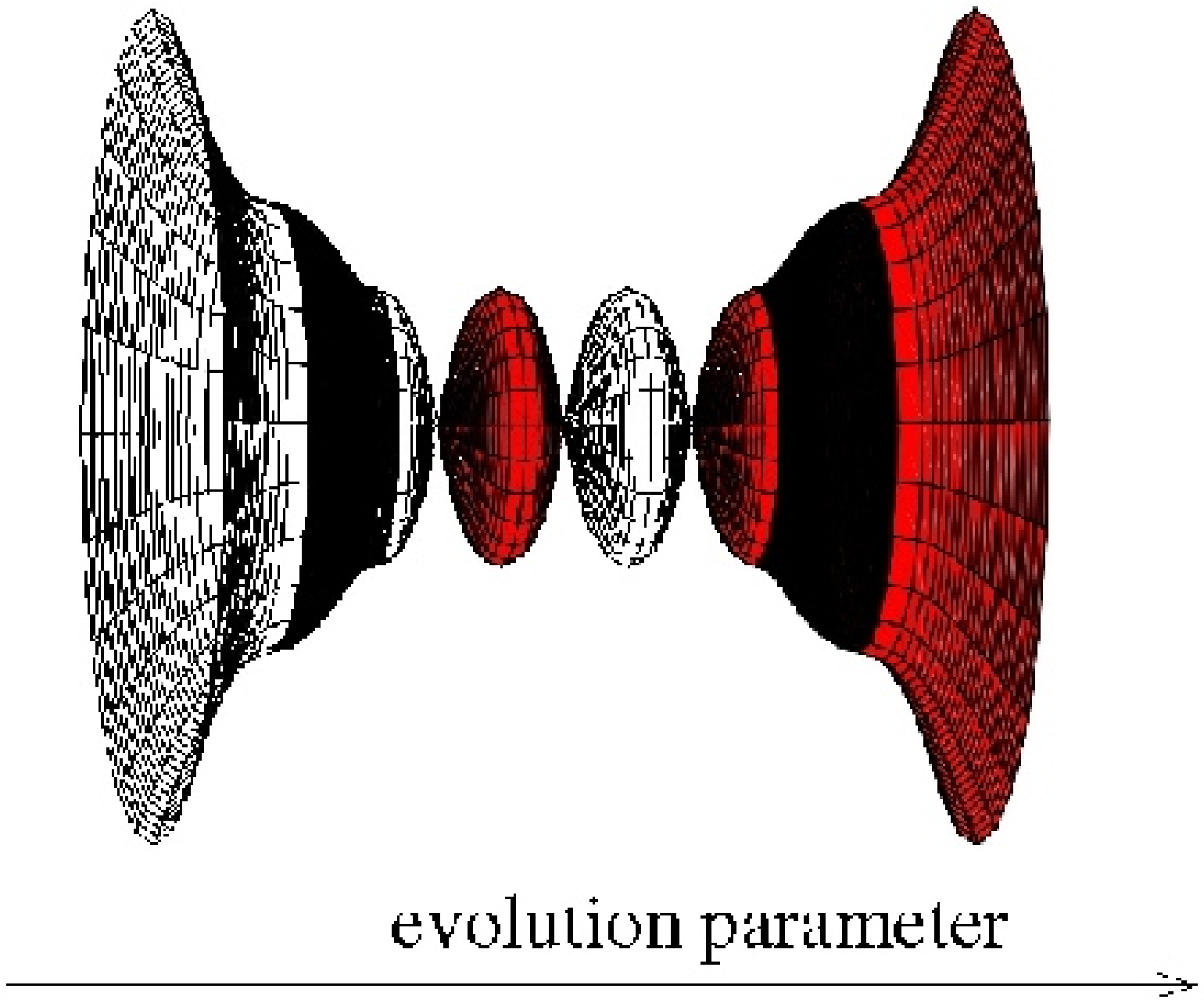}}
\caption{A typical configuration, or ``path'' that contributes to the path integral for resonant tunnelling in the thin wall limit. The lefthand plot shows the evolution of the bubble radius, $R(\la)$. The evolution parameter, $\la$, is given by real time, $t$, in the classically allowed regions, I, III, and V, and by Euclidean time, $\tau$, in the classically forbidden regions II and IV. The transition to negative $R$ should be understood as the vacuum changing inside the bubble. This is shown more intuitively in the righthand plot, which shows the evolution of a circular bubble wall, of positive radius. Colours (red and white) are used  to indicate the interior vacuum in the classically allowed regions, whereas the classically forbidden regions  are coloured in black.} \label{fig:thinwall}}

A typical tunnelling path is shown in Fig.~\ref{fig:thinwall}. The lefthand plot shows the bubble radius as a function of the evolution parameter. The initial state corresponds to a bubble whose radius decreases to a minimum value (region I) before tunnelling along a MPEP (region II) to the oscillatory bound state (region III). At first glance, our plot in region III would not seem to make sense, since the bubble radius clearly appears to go negative. However, we should understand the transition to negative $R$ as a change of vacuum inside the bubble. In other words, the bound state corresponds to a bubble of one vacuum shrinking to zero size, followed by an expanding bubble of different vacuum. The new bubble expands to a maximum radius before contracting again and the process gets repeated. This behaviour is perhaps shown more intuitively in the righthand plot, where the bubble radius is always positive and colours (red and white) are used to indicate the interior vacuum. For a given path, the oscillations in region III can go on any number of times before we tunnel out along another MPEP (region IV), to the final state (region V). As $R$ has changed sign, the final state  corresponds to an eternally expanding bubble of different vacuum to the initial state. This is demonstrated by an overall transition from white to red in the righthand plot.

As mentioned earlier, the governing dynamics is qualitatively the same as in the previous section. We therefore extend the results  to apply to {\it any} quantum field theory, whenever the thin wall limit approximation is valid. Of course, we might question the validity of this approximation for the intermediate bound state, as $R \to 0$. Nevertheless, we have certainly seen in this section and the last that resonant decay of inhomogeneous states is allowed in quantum field theory. 

\section{Conclusions} 
\lab{sec:conc}

In this paper we have reconsidered the issue of resonant tunnelling in scalar quantum field theory, making use of our original  no-go theorem~\cite{Copeland:2007qf} in order to guide us towards an explicit realization of this phenomena. The key development has been to justifiably relax one of the five conditions that led to our theorem, by allowing for solutions with non-zero energy relative to the false vacuum. This permits the existence of finite energy oscillons corresponding to the intermediate bound state, acting as the springboard for the tunnelling amplitude between the initial and final states. Of course, by energy conservation, the initial state will also have non-zero energy, and since it asymptotes to the false vacuum, it must also be inhomogeneous. We found that whilst the homogeneous false vacuum {\it cannot} decay via resonant tunnelling \cite{Copeland:2007qf}, the same is not necessarily true for an {\it in}homogeneous initial state that is merely asymptotically false vacuum.

Although not usually considered in the literature~\cite{Coleman:1977py}, tunnelling from inhomogeneous states is certainly of interest cosmologically. Indeed, the Universe is not expected to be particularly homogeneous or isotropic at early times, before some mechanism flushed away all inhomogeneity and anisotropy on large scales (see, for example~\cite{flush}). That is not to say that it is easy to find an inhomogeneous initial state that will decay by resonant tunnelling. On the contrary, there are a number of obstacles to be overcome if this
is to be realized in practise. In this paper we have considered initial states that correspond to contracting spherical bubbles, with some minimum radius. This is hardly a ``natural'' field configuration, although we note that in principle the initial state can be any non-zero energy solution that asymptotes to the false vacuum and is everywhere stationary at some point in the future. In addition to this rather special initial state, our model must admit localized periodic solutions that act as the intermediate bound state. We must also be able to join this solution at either end  to the initial and final states via Euclidean evolution, connnecting one {\it everywhere-stationary} profile to another. Once all these conditions are satisfied, we must fine tune the energy so that the oscillon action satisfies (\ref{eq:resTunCondQFT}). Then the initial state will decay to the final state with almost unit probability. The final state corresponds to an expanding bubble, whose interior vacuum differs from the interior of the initial state. 

The conditions for resonance have all been derived within the semi-classical approximation. With the additional assumption of spherical symmetry, we have shown that resonant decay of inhomogeneous states is allowed in quantum field theory. It is natural to ask what happens if we increase the number of degrees of freedom by allowing for  non-spherical field configurations. For a spherically symmetric initial state, we do not expect {\it non}-spherically symmetric states to play much of a role, even if we were able to match a significant number of them on to the initial  state via Euclidean evolution. This is because non-spherically symmetric  solutions are typically suppressed relative to the spherically symmetric ones~\cite{minaction}. Even though some non-spherically symmetric oscillons have been found, they rapidly decay into a spherical profile~\cite{mark}.

We must also consider oscillons with a finite lifetime, and the impact they might have on our results. Of course, these are not strictly
periodic but can live for many oscillations, and will contribute to tunnelling paths in the path integral. Given that the Q-factor of a resonant cavity increases
with increasing ``dwell time'' (the length of time that an electron spends in the central well) we may expect
that the finite lifetime of oscillons would tend to broaden the resonance. Another effect which proves
deleterious to resonance is non-zero temperature, again with the consequence of broadening the resonances.

Unfortunately, gravity will also have a negative impact on resonant tunnelling, mainly due to the effect of Hubble damping. Hubble damping eliminates the existence of periodic oscillons, and, given the reasons outlined in the previous paragraph, this can only lead to further broadening of the resonances. One is also faced with the possibilty of gravitational collapse, which could potentially eliminate the  turning point for the collapsing spherical bubble, making tunnelling impossible.

Despite our rather gloomy discussion, we believe our results could have a positive impact on Tye's approach to tunnelling within the string landscape~\cite{Tye:2006tg}. Recall that Tye has argued that  should resonant (or ``fast'') tunnelling occur within the landscape, then we would naturally find ourselves in a low energy vacuum, having repeatedly tunnelled through the landscape along resonant paths.  Crucially, resonant tunnelling has now  been explicitly shown to occur in quantum field theory, albeit only for spherically symmetric, inhomogeneous states, lending some credence to Tye's arguments.

It is fun to speculate as to how the tunnelling process described in this paper could be extended to apply to the landscape. After one resonant decay, the state corresponds to an expanding bubble, enveloping the false vacuum. Globally, this solution is never again stationary and is immune to further decay. Even if we redefine the false vacuum and restrict attention to the bubble's interior, we fall victim to our original no-go theorem~\cite{Copeland:2007qf} and future {\it resonant} tunnelling is impossible. This would seem to be make the cascade of resonant decays desired within the landscape impossible. However, what if the final state after the first decay were not an eternally expanding bubble, but another oscillon, or a subcritical bubble with some maximum size? Such a final state could potentially decay via resonant tunnelling to yet another state. This may seem rather contrived, but this is not a major concern for the landscape, given the huge range of possible configurations available. We could ultimately develop a picture of cascading oscillons within the landscape, with each subsequent oscillon probing yet more vacua!

\end{document}